\documentclass[fleqn,twoside]{article}

\def\noi{\noindent}

\newcommand{\heads}[2]{\markboth{\protect\small\it #1}{\protect\small\it #2}}
\newcommand{\Acknow}[1]{\subsection*{\normalsize Acknowledgement} #1}

\newcommand{\Arthead}[5]{ \setcounter{page}{#4}\thispagestyle{empty}\noi
    \unitlength=1pt \begin{picture}(500,40)
        \put(0,58){\shortstack[l]{\small\it Gravitation \& Cosmology,
                   \small\rm Vol. #1 (#2), No. #3, pp. #4--#5 \\
\footnotesize Issue dedicated to the centennary of Prof. D.D.Ivanenko \\
        \footnotesize \copyright \ #2 \ Russian Gravitational Society}}
    \end{picture}}

\newcommand{\Title}[1]{\noi {\Large #1} \\}
\newcommand{\Author}[2]{\noi {\large\bf #1}\\[2ex]\noindent{\it #2} \\}

\newcommand{\Rec}[1]{\noi {\it Received #1} \\}

\newcommand{\Abstract}[1]{\vskip 2mm \begin{center}
        \parbox{16.4cm}{\small\noi #1} \end{center}\medskip}

\newcommand{\foom}[1]{\protect\footnotemark[#1]}

\newcommand{\email}[2]{\footnotetext[#1]{e-mail: #2}}

\def\nqq{\hspace*{-2em}}

\def\lal{&&\nqq {}}

\def\beq{\begin{equation}}
\def\eeq{\end{equation}}
\def\bear{\begin{eqnarray}}
\def\bearr{\begin{eqnarray} \lal}
\def\ear{\end{eqnarray}}
\def\earn{\nonumber \end{eqnarray}}

\def\yy{\\[5pt] {}}

\def\thebibliography#1{\section*{\normalsize References}
\list
{[\arabic{enumi}]}{\settowidth\labelwidth{[#1]}\leftmargin\labelwidth
\advance\leftmargin\labelsep
\usecounter{enumi}}
\def\newblock{\hskip .11em plus .33em minus -.07em}
\sloppy
\sfcode`\.=1000\relax}

\heads{Alexander Balakin and Veronika Kurbanova}
      {Anomalous Polarization-Curvature Interaction in a Gravitational-Wave Field}

\begin{document}

\twocolumn[
\Arthead{10}{2004}{1-2 (37-38)}{98}{106}

\Title{ANOMALOUS POLARIZATION-CURVATURE INTERACTION\yy
       IN A GRAVITATIONAL-WAVE FIELD}

   \Author{Alexander Balakin\foom 1
           and Veronika Kurbanova\foom 2}
         {Dept. of Relativity and Gravitation Theory,
           Kazan State University, 18 Kremlevskaya St., Kazan 420008, Russia}

\Rec{30 January 2004}

\Abstract
    {An exact solution to the dynamic equations for a massive
     boson traveling in a pp-wave gravitational background under the
     influence of the force induced by curvature, is presented. We
     focus on the effect of anomalous polarization-curvature
     interaction and consider models in which the coupling constant
     of such an interaction is treated to be either a deterministic
     quantity or a random variable.}
]

\email 1 {Alexander.Balakin@ksu.ru}
\email 2 {Veronika.Kurbanova@ksu.ru}

\section{\normalsize Introduction}

The motion of a point particle with intrinsic structure
has become a subject of detailed studies about a century ago.
M.~Abraham~\cite{Abr03} was the first to consider in detail
the motion of an electron as a particle possessing a
supplementary vector degree of freedom. In the twenties of the
past century L.H.~Thomas~\cite{Thom26}, J.~Frenkel~\cite{Fre26}
and I.~Tamm~\cite{Tamm29} studied different aspects of the
classical spinning particle motion. The equations of motion for
particles with arbitrary multipole moments in the framework of
General Relativity were obtained by
M.~Mathisson~\cite{Mat37}  and then revived by
A.~Papapetrou~\cite{Pap51} and W.G.~Dixon~\cite{Dix64}. Studying
the effect of spin precession in an external electromagnetic field in
the framework of special relativity, V.~Bargmann, L.~Michel and
V.L.~Telegdi~\cite{BMT59} introduced the term ``anomalous
magnetic moment of an electron''. In the absence of an anomalous
moment, the Bargmann-Michel-Telegdi (BMT) equations coincide with
Bloch's equations~\cite{Blo46}. In the last three decades, the
interest in studying the dynamic equations for particles with
intrinsic degrees of freedom, being derived in the framework of
the Lagrangian or Hamiltonian formalism, has grown considerably
[10-33]. Thereupon, the BMT-model has gained the status of a test
model for verification of new models of dynamics of a point
particle with spin or polarization in external fields of various
kinds. The history of studying the dynamics of spinning particles is
presented in the book~\cite{BBB-TRRP}; alternative approaches
to derivation of motion equations for charge and spin, the
problem of non-collinearity of momentum and velocity four-vectors
and the origin and interpretation of the Thomas precession are
examined in detail there.

The most important thing while analyzing the dynamic equations for
particles with intrinsic structure in the gravitational field is
the studying of a spin-curvature (polarization-curvature)
interaction. Per se, the Mathis\-son-Papapetrou-Dixon equations are a
non-minimal generalization of the classical dynamic equations
because they take into consideration the Riemann curvature tensor
and its covariant derivatives. Formally, the introduction of
curvature-dependent forces (or ``tidal'' forces by the conventional
terminology) breaks the free fall universality principle and thus
contradicts Einstein's equivalence principle. Thereupon, the
spin (polarization) -curvature interaction is now considered to be
a theoretical prediction whose experimental verification may have
an interesting application to the problem of testing the metric
theories of gravity. Following I.B.~Khriplovich
\cite{Khr89,KhP96,PSKh2000,KhP2000} and M.~Bander and
K.~Yee~\cite{BY93}, using an analogy with the anomalous interaction
of a spinning particle and the electromagnetic field, introduced
in the BMT theory, it seems reasonable to consider also the anomalous
interaction of spin (polarization) with the gravitational field.
Such an analysis involves a new parameter $Q$ as a constant of
anomalous spin (polarization)-curvature interaction, by analogy with
the phenomenological parameter $g-2$, introduced in
the BMT theory~\cite{BMT59} as a constant of anomalous interaction of
spinning particle with the electromagnetic field. Under a generalization of
such kind, spin (polarization) dynamics is predetermined by a
Lorentz-type force, one addend of which is proportional to the
Riemann curvature tensor~\cite{BY93}.

In this paper we suggest an exactly integrable model of the dynamics
of a relativistic particle with an arbitrary directed polarization
vector in the gravitational radiation field. The model is
based on the known dynamic equations for particles with
supplementary degrees of freedom and takes into consideration
an anomalous interaction of polarization with curvature. The paper
develops our recent
works~\cite{BK_FF_2001,BK_SS_2001,BKZ2002,BKZ2003}, in which, in
particular, the effect of parametric oscillations of Faraday
rotation of the polarization vector was predicted. The results
obtained in this paper demonstrate a possibility of new type for
the precession of the polarization vector, indicated as
``hyperbolic rotation". We consider two types of models:
first, a model with a deterministic coupling constant of
anomalous interaction between the space-time curvature and particle
polarization, and, second, the model with a random coupling constant.
One of the purposes of this paper is to compare the behaviour of
particle in two different cases: first, when the anomalous
polarization-curvature interaction is absent (the coupling
constant is zero identically), second, when the random
coupling constant is zero {\it on the average}.

The paper is organized as follows. In Sec.~2,
evolution equations for the velocity four-vector and the spin
(polarization) four-vector are established and compared with the
known dynamic equations for particles with an intrinsic structure.
In Sec.~3, exact solutions to the resulting evolution
equations are found for the case when a boson travels in a
pp-wave gravitational background under the influence of
a deterministic force induced by curvature. In Sec.~4, we
discuss the specific features of boson motion in the case when the
coupling constant of the polarization-curvature interaction is
considered as a random variable with zero mean value. The last
section contains conclusions.

\medskip

\section{\normalsize Evolution equations}

\subsection{\small General model}

It is well-known that, for a point particle with supplementary
tensor degrees of freedom, the generalized momentum four-vector
$P^i$ is not parallel to the velocity four-vector
$U^i\equiv\frac{dx^i}{d\tau}$ (see, e.g., \cite{BBB-TRRP})
\begin{equation}\label{1}
P^i = mcU^i+q^i \,,\quad q^i U_i = 0 \,.
\end{equation}
The part of the momentum $q^i$, which is orthogonal to the
velocity four-vector, may include the angular momentum tensor
$S^{ij}$, the quadrupole momentum tensor $J^{imns}$ as well as the
tensors describing higher multipole moments~\cite{Mat37,Gem97-2}
\begin{equation}\label{2}
q^i\equiv U_j \dot{S}^{ij} -\frac43R_{nms \cdot}^{ \ \ \ \
[j}J^{i]mns} U_j+ ...\,.
\end{equation}
The dot denotes the derivative $D/D\tau=U^i\nabla_i$,
where $\nabla_i$ is the covariant derivative. The Riemann
curvature tensor $R_{nmsj}$ and its covariant derivatives happen
to be included into the formulae for the generalized momentum as
necessary structure elements. As usual, the scalar product
$U_iU^i={\rm const}$ is set equal to unity identically,
\begin{equation}\label{3}
U_i U^i=1 \,.
\end{equation}
The scalar product $P^i P_i=m^2c^2+q^iq_i$ does not remain
constant along the particle word-line, or, in other words, it is
not an integral of motion. In our opinion, it is then more
convenient to construct the dynamic model for particles with
intrinsic structure using the velocity four-vector then. Starting
from the relation
\begin{equation}\label{4}
U^i\dot{U}_i=\frac12\frac{D}{D\tau}(U^iU_i)=0 \,,
\end{equation}
one can find the generic form of an evolution equation for the
velocity four-vector:
\begin{equation}\label{5}
\dot{U}^i=f^i\,,\quad f^iU_i=0\,.
\end{equation}
Using the identity:
\begin{equation}\label{6}
f_i \equiv U_i (f_k U^k) + \Delta_i^k f_k \,, \quad \Delta^k_i
\equiv\delta^k_i-U_iU^k \,,
\end{equation}
and the orthogonality condition $f_kU^k=0$, one can represent the
effective force four-vector $f^i$ in the form
\begin{equation}\label{7}
f^i= \Delta_i^k  \ V_k(x,U,S) \equiv (V_i U_k- U_i V_k) U^k \,.
\end{equation}
Here $V^k(x,U,S)$ is an arbitrary vector function depending on
the coordinates, the velocity four-vector and spin (polarization)
variables. Introducing the new antisymmetric tensor (simple
bivector)
\begin{equation}\label{8}
\Omega_{ik}(x,U,S) \equiv V_iU_k-U_iV_k \,,
\end{equation}
one can represent the dynamic equation (\ref{5}) in two
equivalent forms
\begin{equation}\label{9}
\dot{U}^i= \Delta_i^k \ V_k(x,U,S) \leftarrow \cdot \rightarrow
\dot{U}_i= \Omega_{ik}(x,U,S) U^k \,.
\end{equation}
To characterize the state of a particle which possesses a vector
degree of freedom (spin or polarization), one uses the well-known
decomposition of the anti-symmetric tensor of total moment
\begin{equation}\label{10}
S_{ik}=L_iU_k-L_kU_i-\epsilon_{iklm}S^lU^m\,.
\end{equation}
Here $L^i$ is the four-vector of orbital moment and $S^i$ is the
spin (polarization) four-vector, defined as follows:
\begin{equation}\label{11}
L_i = S_{ik} U^k \,, \quad
S_i\equiv\frac12\epsilon_{iklm}S^{lm}U^k\equiv S^{*}_{ik}U^k \,,
\end{equation}
where $\epsilon_{iklm}$ is the Levi-Civita tensor
\begin{equation}\label{12}
\epsilon_{iklm} = \sqrt{-g} E_{iklm} \,, \quad E_{0123}= -1 \,,
\end{equation}
and the asterisk in (\ref{11}) introduces the corresponding
duality symbol. The four-vectors $L_i$ and $S_i$ are orthogonal
to the velocity four-vector automatically by the definition
(\ref{11})
\begin{equation}\label{13}
L^i U_i = 0 \,,  \quad S^{k}U_k = 0 \,.
\end{equation}
In this sense the so-called Tamm condition $S^{k}U_k = 0$ is not
an additional requirement. One can find an obvious analogy
between the decomposition (\ref{11}) and the decomposition of the
Maxwell tensor $F_{ik}$ in electrodynamics~\cite{Maugin}
\begin{equation}\label{14}
F_{ik}=E_i U_k - E_k U_i - \epsilon_{iklm} B^l U^m\,.
\end{equation}
Following this analogy, one can say, that $S^i$
plays the same role as the four-vector of magnetic induction $B^i$
(and is in fact a pseudo-vector), while $L^i$
plays the role of electric field four-vector $E^i$ (and is a true
vector). The four-vector $S^i$ is considered to be a vector
parameter of the intrinsic state of a particle. The scalar square
of this four-vector is assumed to be constant along the particle
world-line, i.e.,
\begin{equation}\label{15}
S^{k}S_k = - E^2 = {\rm const} \neq 0 \,.
\end{equation}
This requirement results in
\begin{equation}\label{16}
S^{k}\dot{S}_k = 0 \,,
\end{equation}
and, following the previous discussion concerning the $\dot{U}^i$
decomposition, one can state that the corresponding evolution
equation has two equivalent forms
\begin{eqnarray}\label{17}
\dot{S}^i= \left(\delta_i^k-\frac{S_iS^k}{S_lS^l}\right)
W_k(x,U,S)  \leftarrow \cdot \rightarrow\nonumber\\
\dot{S}_i=\omega_{ik}(x,U,S) S^k \,,
\end{eqnarray}
where
\begin{equation}\label{18}
\omega_{ik} \equiv \frac{W_iS_k-S_iW_k}{S^lS_l} \,.
\end{equation}
From the orthogonality condition $S_i U^i=0$ it follows that
\begin{equation}\label{19}
U^i\dot{S}_i+S^i\dot{U}_i=0\,,
\end{equation}
consequently, the four-vectors $V_i$ and $W_i$ as well as
the tensors $\Omega_{ik}$ and $\omega_{ik}$ are connected by the
requirements
\begin{equation}\label{20}
U^i W_i + S^i V_i = 0 \,, \quad \left(\omega_{ik} -
\Omega_{ik}\right) U^iS^k =0  \,.
\end{equation}
This means, in particular, that
\begin{equation}\label{21}
\omega_{ik} S^k = \Omega_{ik} S^k +
\widetilde{\Omega}_{ik}(x,U,S) U^k\,,
\end{equation}
with
\begin{equation}\label{22}
\widetilde{\Omega}_{ik}=-\widetilde{\Omega}_{ki},\quad
\widetilde{\Omega}_{ik}U^iS^k=0 \,.
\end{equation}
Thus, we can represent the evolution equations of a particle with
spin or polarization in the following generic form:
\begin{equation}\label{23}
\dot{U}_i = \Omega_{ik} U^k \,, \quad \dot{S}_i = \Omega_{ik} S^k
+ \widetilde{\Omega}_{ik}U^k \,.
\end{equation}

\subsection{\small Comparison with well-known models}
\subsubsection{The Bargmann-Michel-Telegdi model}

To recover the BMT results \cite{BMT59}
\begin{eqnarray}\label{24}
\dot{U}_i=\frac{e}{mc^2}F_{ik}U^k\,,\nonumber\\
\dot{S}_i=\frac{e}{mc^2}\left[\frac g2F_{ik}S^k+U_i \left(\frac
g2-1\right)F_{kl}S^kU^l\right] \,,
\end{eqnarray}
where $g$ is a gyro-magnetic ratio, one has to suppose that
$\Omega_{ik}$  is proportional to the Maxwell tensor $F_{ik}$:
\begin{equation}\label{25}
\Omega_{ik} = \frac{e}{mc^2} F_{ik} \,,
\end{equation}
and $\widetilde{\Omega}_{ik}$ has to be equal to
\begin{equation}\label{26}
\widetilde{\Omega}_{ik} = \frac{e}{mc^2}\left(\frac g2-1\right) (
U_{k}F_{il} - U_{i}F_{kl})S^l \,.
\end{equation}

\subsubsection{The Bander - Yee model}

Bander and Yee \cite{BY93} obtained the following spin evolution equation:
\begin{eqnarray}\label{27}\nonumber
\frac{DS_{ab}}{D\tau}+(U_bS_{ac}-U_aS_{bc})\frac{DU^c}{D\tau}
\\=
-\left( \frac{eg}{4mc^2}F^{cd}+\frac\kappa{8mc}R^{cdef}S_{ef}
\right)\nonumber
\\\times(S_{ac}\Delta_{bd}+S_{bd}\Delta_{ac})\,.
\end{eqnarray}
The dynamic equations derived in \cite{BY93} have terms which
are second-order in the spin, terms with derivatives
of the Maxwell tensor and a term with the second order derivative
of $U^i$. In the approximation where we neglect all these terms,
we have the following equation~\cite{BY93}:
\begin{equation}\label{28}
\dot{U}_i=\left[\frac{e}{mc^2}F_{ik}+\frac1{2mc}R_{ikab}S^{ab}\right]U^k\,.
\end{equation}
Let us consider a model with vanishing orbital moment ($L_i
\equiv 0$). Then, in terms of the spin (polarization) four-vector
$S^a$,
the dynamic equations (\ref{28}) have the form (\ref{23}) if
\begin{equation}\label{29}
\Omega_{ik}=\left[\frac{e}{mc^2}F_{ik}-\frac1{mc}R^*_{ikab}S^aU^b\right]
\,,
\end{equation}
where
\begin{equation}\label{30}
R^*_{ikab}\equiv\frac12 R_{ik}^{\ \ cd} \epsilon_{cdab}
\end{equation}
is the right-dual Riemann tensor. This quantity is a
pseudo-tensor, but the product $R^*_{ikab}S^a$ happens to be a
true tensor. Eqs. (\ref{27}) can be represented in the
form (\ref{23}) when
\begin{equation}\label{31}
\widetilde{\Omega}_{ik}= \left[U_{k}  \widehat{\Omega}_{il} -
U_{i} \widehat{\Omega}_{kl} \right]S^l \,,
\end{equation}
with
\begin{equation}\label{32}
\widehat{\Omega}_{ik}\equiv \frac{e}{mc^2}F_{ik}\left(\frac
g2-1\right)- \frac{\kappa-1}{mc}R^*_{ikab}S^aU^b \,.
\end{equation}

\subsection{\small Equations of boson dynamics}

Let us consider the dynamics of vector bosons possessing zero
charge $e=0$ and polarization four-vector $S^i$. The dynamic
equation for such a particle can be immediately obtained from
(\ref{28}):
\begin{equation}\label{33}
\dot{U}_i=-\frac1{mc}R^*_{iklm}U^kS^lU^m\,.
\end{equation}
The equation of polarization evolution, obtained from (\ref{23})
with (\ref{29}), (\ref{31}) and (\ref{32}), yields
\begin{eqnarray}\label{34}
\dot{S}_i=-\frac1{mc} \left[ \kappa R^*_{iklm}S^kS^lU^m
\right.\nonumber\\+\left.
(\kappa-1)U_iR^*_{klmn}S^kU^lS^mU^n \right].
\end{eqnarray}
Eqs. (\ref{33}) and (\ref{34}) can be also obtained from
the BMT equations by a formal substitution discussed by
Khriplo\-vich in \cite{KhP2000}:
\begin{equation}\label{35}
\frac{e}{c} F_{ik}\quad \longrightarrow\quad \frac{1}{2}
R_{ikab}S^{ab} = - R^*_{ikab} S^a U^b \,,
\end{equation}
if one replaces $g/2$ by $\kappa$. One sees that the case
$\kappa \equiv 1$ corresponds to the BMT model without an anomalous
magnetic moment. We introduce a new dimensionless parameter $Q$
by the definition $Q=\kappa -1$ and suppose that, from a
phenomenological point of view, the $Q$ parameter is a coupling
constant describing an anomalous interaction of
polarization with curvature. Finally, one obtains the following
equation of evolution for the polarization four-vector:
\begin{eqnarray}\label{36}
\dot{S}_i=-\frac{1}{mc} \left[ (Q+1)\cdot R^*_{iklm}S^kS^lU^m
\right.\nonumber\\\left.+
Q\cdot U_iR^*_{klmn}S^kU^lS^mU^n \right] \,.
\end{eqnarray}

\section{\normalsize Boson dynamics in a pp-wave gravitational field. Deterministic model}

\subsection{\small Gravitational wave background}\label{GR}

The well-known pp-wave solution of the vacuum Einstein equations
can be represented by the metric \cite{KSHM-ESEE}
\begin{eqnarray}\label{37}
d\tau^2=2dudv-L^2 \left\{\left[{\rm e}^{2\beta}(dx^2)^2
\right.\right.
\nonumber\\
+\left.\left.
{\rm e}^{-2\beta} (dx^3)^2 \right]
\cosh2\gamma+2\sinh2\gamma dx^2dx^3\right\},
\end{eqnarray}
where
\begin{equation}\label{38}
u\equiv \frac{1}{\sqrt{2}}(ct-x^1),\qquad
v\equiv \frac{1}{\sqrt{2}}(ct+x^1)
\end{equation}
are the retarded and advanced time, respectively. The
functions $\beta(u)$ and $\gamma(u)$ are considered to be
arbitrary functions of the retarded time, satisfying the
following conditions on the wave front plane $u=0$:
\begin{equation}\label{39}
\beta(0)=\gamma(0)=0 \,,  \quad \beta{}'(0)=\gamma{}'(0)=0 \,.
\end{equation}
The background factor $L(u)$ satisfies the equation
\begin{equation}\label{40}
L''+L\left(\beta'{}^2\cosh^2{2\gamma}+\gamma'{}^2\right)=0
\end{equation}
and the initial data
\begin{equation}\label{41}
L(0)=1\,,\quad L{}'(0)=0 \,.
\end{equation}
The gravitational wave (GW) is indicated as a GW with the first
polarization when $\beta(u) \neq 0$, $\gamma(u)=0$, and as a GW
with the second polarization when $\beta(u) = 0$, $\gamma(u)
\neq 0$. For the pp-wave solution~(\ref{37}), the right-dual
Riemann tensors $R^*_{jklm}$~(\ref{30}) and the left-dual one
$^*R_{jklm}\equiv\frac 12 \epsilon^{\ \
pq}_{jk\cdot\cdot}R_{pqlm}$ coincide and have the following
nonvanishing components:
\begin{eqnarray}\label{42}
R^*_{2u2u} = -\left[L^2{\rm e}^{
2\beta}\left(\gamma' - \frac12\beta'\sinh4\gamma\right)\right]',
\nonumber\\
R^*_{3u3u} = \left[L^2{\rm e}^{-2\beta}\left(\gamma' +
\frac12\beta'\sinh4\gamma\right)\right]',
\nonumber\\
R^*_{2u3u}=R^*_{3u2u} = \left[L^2\beta'\cosh^2 2\gamma\right]'.
\end{eqnarray}
Here and in what follows, Latin indices run over four values:
$u,v,2,3$, while Greek indices run over two values: $2,3$.

\subsection{\small Exact solutions to the evolution equations}

\subsubsection{Key subsystem of the evolution equations}

The GW metric~(\ref{37}) admits two spacelike Killing vectors and
one null covariantly constant Killing vector \cite{KSHM-ESEE}
\begin{eqnarray}\label{43}
\mathop{\xi^i}\limits_{(2)} = \delta^i_2 \,, \quad
\mathop{\xi^i}\limits_{(3)} = \delta^i_3 \,, \quad
\mathop{\xi^i}\limits_{(v)} = \delta^i_v \,, \\
g_{ik}\mathop{\xi^i}\limits_{(v)} \mathop{\xi^k}\limits_{(v)} = 0 \,,
\quad \nabla_k \mathop{\xi^i}\limits_{(v)}  = 0 \,.
\end{eqnarray}
The projection of the dynamic equation (\ref{33}) onto the
direction given by $\mathop{\xi^i}\limits_{(v)}$ yields
\begin{equation}\label{44}
\frac{d U_v}{d\tau} = 0 \rightarrow U_v =  {\rm const} \equiv C_v.
\end{equation}
It follows from (\ref{44}) that the parameter $\tau$ is linearly
connected with the $u$ coordinate since
\begin{equation}\label{45}
\frac{du}{d\tau} \equiv U^u=U_v=C_v \,.
\end{equation}
When $C_v \neq 0$ (this condition does not hold only for a massless
particle co-moving the GW), one obtains
\begin{equation}\label{46}
\tau=\tau_0+\frac{u}{C_v} \,.
\end{equation}
Since $U^iU_i =1$, the $U_u$ component of the velocity four-vector
can be found immediately:
\begin{equation}\label{47}
U_u=\frac1{2C_v}\left[1-g^{\alpha\sigma}U_\alpha U_\sigma\right]
\,.
\end{equation}
Similarly, from $S^i U_i = 0$, one obtains an expression for
the $S_u$ component of the polarization four-vector:
\begin{equation}\label{48}
S_u=-\frac1{C_v}\left[S_vU_u+g^{\alpha \sigma}U_\alpha
S_\sigma\right] \,.
\end{equation}
Thus the $U_u$ and $S_u$ quantities are found as soon as $U_2$,
$U_3$, $S_2$, $S_3$ and $S_v$ become known. For the remaining five
components of the velocity and polarization four-vectors, one has five
equations:
\begin{eqnarray}\label{49}
\frac{dU_\alpha}{du}=-\frac 1{mc}R^*_{\alpha u \sigma u}
\left(C_v S^\sigma-S_vU^\sigma\right) \,,
\nonumber\\
C_v\frac{dS_\alpha}{du} =
\left(C_v S^\sigma-S_vU^\sigma\right)
\nonumber\\
\times
\left[\frac{g'_{\alpha\sigma}}2-
\frac{Q+1}{mc}S_vR^*_{\alpha u \sigma u}
\right.
\nonumber\\
\left. -\frac{Q}{mc}U_\alpha R^*_{\rho u \sigma u} \left(C_v
S^\rho-S_vU^\rho\right) \right] \,,
\nonumber\\
\frac{dS_v}{du}=-\frac Q{mc} R^*_{\alpha u \sigma u}
\left(C_v S^\alpha-S_v U^\alpha\right)
\nonumber\\
\times
\left(C_v S^\sigma-S_v U^\sigma\right)
\end{eqnarray}
with the following initial conditions:
\begin{equation}\label{50}
U_\alpha(0)=C_\alpha,\quad S_\alpha(0)=E_\alpha,\quad S_v(0)=E_v
\,.
\end{equation}
After introducing the new variables
\begin{equation}\label{51}
X_\alpha\equiv C_vS_\alpha-S_vU_\alpha
\end{equation}
a closed subset of three equations emerges from the
set~(\ref{49})
\begin{equation}\label{52}
\frac{d X_\alpha}{du} =\left[a_\alpha^\sigma(u)- \frac{Q}{mc}S_v
R^*{}^\sigma_{\cdot u \alpha u}(u)\right]X_\sigma \,,
\end{equation}
\begin{equation}\label{53}
\frac{dS_v}{du}=-\frac Q{mc} R^*_{\alpha u \sigma u}(u)X^\alpha
X^\sigma \,,
\end{equation}
where
\begin{equation}\label{54}
a_\alpha^\sigma(u)\equiv \frac 12
g^{\sigma\rho}(u)g'{}_{\rho\alpha}(u).
\end{equation}
The prime denotes a derivative with respect to the retarded
time $u$. For the transversal components of the velocity
four-vector $U_\alpha$ we obtain two decoupled equations
\begin{equation}\label{55}
\frac{dU_\alpha}{du}=-\frac 1{mc}R^*_{\alpha u \sigma u}X^\sigma
\,,
\end{equation}
which give the following formal solutions for $U_\alpha$:
\begin{equation}\label{56}
U_\alpha(u)=C_\alpha- \frac 1{mc}\int\limits_0^u R^*_{\alpha u
\sigma u}(\tilde u)X^\sigma(\tilde u)\ d\tilde u \,.
\end{equation}
The unknowns $S_\alpha(u)$ can be found immediately from the
relation (\ref{51}):
\begin{equation}\label{57}
S_\alpha(u)=\frac1{C_v}\left[X_\alpha(u)+S_vU_\alpha(u)\right] \,.
\end{equation}
Thus all the solutions of Eqs. (\ref{49}) will be
represented by quadratures as soon as solutions to the system
(\ref{52})-(\ref{53}) are known. We will designate the latter as
{\em the key subsystem}.

\subsection{\small Exact solutions to the key subsystem for the case $Q\equiv 0$}

One of the purposes of the paper is to compare the behaviour of
a vector boson in the GW background in two cases. The first one
takes place when the coupling constant $Q$ is zero
identically, i.e., when there is no anomalous interaction of
polarization with curvature. The second case is connected with
the suggestion that $Q$ effectively depends on the environment of
the boson and is a random variable with zero mean value. Thus we
first consider the evolution equations with $Q \equiv 0$.
From~(\ref{53}) it directly follows  that
\begin{equation}\label{58}
S_v(u)={\rm const}=S_v(0)\equiv E_v \,,
\end{equation}
and Eqs.~(\ref{52}) result in
\begin{equation}\label{59}
X'_\alpha =a_\alpha^\sigma(u) X_\sigma \,.
\end{equation}
Such a system have been solved in~\cite{Bal97} and discussed
in [35-38,41,42], and here we use the result:
\begin{eqnarray}\label{60}
\left(\begin{array}{c} X_2(u) \\ X_3(u) \end{array}\right)
\nonumber\\
=
L
\left(\begin{array}{cc}{\rm e}^{\beta} & 0 \\ 0 & {\rm e}^{-
\beta}\end{array}\right)
\left(\begin{array}{cc}
\cosh\gamma & \sinh\gamma \\ \sinh\gamma & \cosh\gamma\end{array}\right)
\nonumber\\
\times
\left(\begin{array}{cc}
\cos\psi & -\sin\psi \\ \sin\psi &  \cos\psi\end{array} \right)
\left(\begin{array}{c} C_vE_2-E_vC_2 \\ C_vE_3-E_vC_3 \end{array} \right),
\end{eqnarray}
where
\begin{equation}\label{61}
\psi(u) \equiv \int\limits_0^u \beta '(\tilde u) \sinh 2\gamma(\tilde u)\
d\tilde u.
\end{equation}
The precession of the polarization four-vector, described by Eqs.
(\ref{57}), (\ref{56}), (\ref{60}), (\ref{61}) was discussed in detail in Ref.
\cite{BKZ2002}.

\subsection{\small Exact solutions to the key subsystem for the case $Q \neq 0$}

To gain analytic progress in the case of an anomalous interaction
of polarization with curvature, let us consider the GW to have
the first polarization ($\gamma = 0$). The right-dual curvature
tensor~$R^*_{klmn}$ (\ref{42}) has only two nonzero components:
\begin{equation}\label{62}
R^*_{2u3u}=R^*_{3u2u}=(L^2\beta')'\,,
\end{equation}
and thus the equations for $X_\alpha$ yield:
\begin{equation}\label{63}
X_2'=\left(\frac{L'}L+\beta'\right)X_2- \frac
Q{mc}S_ve^{2\beta}\frac{(L^2\beta')'}{L^2}X_3 \,,
\end{equation}
\begin{equation}\label{64}
X_3'=\left(\frac{L'}L-\beta'\right)X_3- \frac
Q{mc}S_ve^{-2\beta}\frac{(L^2\beta')'}{L^2}X_2 \,.
\end{equation}
The substitution
\begin{equation}\label{65}
X_2=L{\rm e}^{\beta} Y_2 \,, \quad X_3=L{\rm e}^{-\beta} Y_3
\end{equation}
results in the following equations for the new variables $Y_2$ and
$Y_3$:
\begin{equation}\label{66}
Y'_2=\Omega(u)Y_3 \,, \quad  Y'_3=\Omega(u)Y_2 \,,
\end{equation}
\begin{equation}\label{67}
\Omega(u)\equiv -\frac Q{mc} \ S_v(u) \
\frac{[L^2(u)\beta'(u)]'}{L^2(u)} \,.
\end{equation}
Solutions to this system can be represented in the elementary
functions:
\begin{eqnarray}\label{68}
Y_2(u)=Y_2(0)\cosh\Psi(u) + Y_3(0)\sinh\Psi(u) \,,
\nonumber\\
Y_3(u)=Y_3(0)\cosh\Psi(u) + Y_2(0)\sinh\Psi(u) \,,
\end{eqnarray}
where
\begin{equation}\label{69}
\Psi\equiv\int_0^u\Omega(\tilde u) \tilde u \,, \qquad
\Psi(0)\equiv 0 \,,
\end{equation}
\begin{eqnarray}\label{70}
Y_2(0) = C_v E_2 - E_v C_2\,, \nonumber\\
Y_3(0) = C_v E_3 - E_v C_3\,.
\end{eqnarray}
As for the unknown $S_v(u)$, using the expressions for quadratic
integrals of motion $U^iU_i{=}1$, $S^iU_i{=}0$,
$S^iS_i{=}{-}E_0^2$, as well as the definition of
$X_\alpha(u)$~(\ref{51}), and the expressions for
$Y_\alpha(u)$~(\ref{68}), one can show that the following
relations take place:
\begin{eqnarray}\label{71}
S^2_v(u)=C_v^2E_0^2-\left[Y_2^2(0)+Y_3^2(0)\right]\cosh2\Psi(u)
\nonumber\\
-2Y_2(0)Y_3(0)\sinh2\Psi(u) \equiv {\cal B}^2(\Psi) \,,
\end{eqnarray}
\begin{equation}\label{72}
E^2_v = C_v^2 E_0^2 - Y^2_2(0) - Y^2_3(0) \,.
\end{equation}
The relations (\ref{67}),(\ref{69}),(\ref{71}) make it
possible to find $\Psi(u)$ by quadratures:
\begin{equation}\label{73}
\pm \int\limits_{0}^{\Psi(u)} \frac{d\widetilde\Psi}{{\cal
B}(\widetilde\Psi)} = -\frac Q{mc} r^{-1}(u) \,,
\end{equation}
where
\begin{equation}\label{74}
r^{-1}(u) \equiv  \int\limits_0^u R^3_{\cdot u3u}(\tilde u)
d\tilde u = \int\limits_0^u \left\{\frac{(L^2\beta')'}{L^2}
\right\}(\tilde u)d\tilde u \,.
\end{equation}
The function $r(u)$ has the dimension of distance; it may be
considered as some effective curvature radius.

Generally, the solution for $\Psi(u)$ can be represented in terms
of elliptic functions. Below we consider some approximations to
such solutions. When the function $\Psi(u)$ is known, the
transversal components of the velocity four-vector and
the polarization four-vector take the form
\begin{eqnarray}\label{75}
U_2(u)=C_2+
 \frac{Y_3(0)}{mc}\int\limits_0^u
\left\{[L^2\beta']'\frac{{\rm e}^{\beta}}{L}\cosh\Psi
\right\}(\tilde u) d\tilde u
+\nonumber\\
+\frac{Y_2(0)}{mc}\int\limits_0^u \left\{[L^2\beta']'\frac{{\rm
e}^{\beta}}{L}\sinh\Psi \right\}(\tilde u) d\tilde u \,,
\end{eqnarray}
\begin{eqnarray}\label{76}
U_3(u)=C_3+
 \frac{Y_2(0)}{mc}\int\limits_0^u
\left\{[L^2\beta']'\frac{{\rm e}^{-\beta}}{L}\cosh\Psi
\right\}(\tilde u) d\tilde u
+\nonumber\\
+\frac{Y_3(0)}{mc}\int\limits_0^u \left\{[L^2\beta']'\frac{{\rm
e}^{-\beta}}{L}\sinh\Psi \right\}(\tilde u) d\tilde u \,,
\end{eqnarray}
\begin{eqnarray}\label{77}
S_2(u) = \frac{Y_2(0)}{C_v}L(u){\rm e}^{ \beta(u)}\cosh\Psi(u)
\nonumber\\
+\frac{Y_3(0)}{C_v}L(u){\rm e}^{ \beta(u)}\sinh\Psi(u)
+S_v(u)\frac{U_2(u)}{C_v} \,,
\end{eqnarray}
\begin{eqnarray}\label{78}
S_3(u) = \frac{Y_3(0)}{C_v}L(u){\rm e}^{-\beta(u)}\cosh\Psi(u)
\nonumber\\
+\frac{Y_2(0)}{C_v}L(u){\rm e}^{-\beta(u)}\sinh\Psi(u)
+S_v(u)\frac{U_3(u)}{C_v} \,.
\end{eqnarray}
Eqs. (\ref{75})-(\ref{78}) display a new type of precession of
the polarization vector, which can be designated as {\it hyperbolic
rotation} of polarization under the influence of an anomalous
curvature force. The above quantities depend non-linearly on
the $Q$ parameter via the function $\Psi(u)$ given by
Eqs. (\ref{73}) and (\ref{71}). When~$Q \equiv 0$,
$\Psi(u)=0$, and the integrals in (\ref{75}) and~(\ref{76}) take
an explicit form:
\begin{eqnarray}\label{79}
U_{2 |Q=0}(u)=C_2
+
\frac1{mc}(C_vE_3-E_vC_3)[L(u){\rm e}^{\beta(u)}]' \,,
\end{eqnarray}
\begin{eqnarray}\label{80}
U_{3 |Q=0}(u)=C_3
-
\frac1{mc}(C_vE_2-E_vC_2)[L(u){\rm e}^{-\beta(u)}]' \,,
\end{eqnarray}
\begin{eqnarray}\label{81}
S_{2 |Q=0}(u)=E_v\frac{C_2}{C_v}
+
\left(E_2-E_v\frac{C_2}{C_v}\right)L(u){\rm e}^{ \beta(u)}
\nonumber\\
+\frac{E_v}{mc}\left(E_3-E_v\frac{C_3}{C_v}\right)[L(u){\rm e}^{
\beta(u)}]' \,,
\end{eqnarray}
\begin{eqnarray}\label{82}
S_{3 |Q=0}(u)=E_v\frac{C_3}{C_v}
+
\left(E_3-E_v\frac{C_3}{C_v}\right)L(u){\rm e}^{-\beta(u)}
\nonumber\\
-\frac{E_v}{mc}\left(E_2-E_v\frac{C_2}{C_v}\right)[L(u){\rm
e}^{-\beta(u)}]'\,.
\end{eqnarray}
Thus Eqs.(\ref{44}), (\ref{47}), (\ref{48}), (\ref{73}),
(\ref{75})-(\ref{78}) describe the evolution of a relativistic
polarized boson in a pp-wave gravitational background in the
framework of a model with the deterministic parameter $Q$ (coupling
constant of the polarization-curvature interaction). Eqs.
(\ref{79})-(\ref{82}), describing the particular case $Q=0$, will
be used directly in the next section.

\section{\normalsize Random polarization-curvature interaction}

We suppose that the anomalous interaction of polarization with
curvature has a random nature and can be attributed to the
particle interaction with its environment mediated by the space-time
curvature. Mathematically, this means that the parameter $Q$ can be
treated as a random variable. Basing on such an ansatz, one can
calculate the mean values of the velocity and polarization
four-vectors and compare the results with those for $Q\equiv 0$,
given by Eqs. (\ref{79})-(\ref{82}). The distribution
function for the $Q$ variable is supposed to be Gaussian:
\begin{equation}\label{83}
f(Q) = \frac1{{\cal D} \sqrt\pi} \exp\left(-\frac{Q^2}{{\cal
D}^2}\right),
\end{equation}
where ${\cal D}$ is a dispersion parameter.

\subsection{\small First special solution: $Y_2(0) = Y_3(0) \equiv 0$}

To obtain a starting exact result, let us suppose that both
initial values $Y_2(0)$ and $Y_3(0)$ are equal to zero. In this
case it follows from (\ref{71})  that $S_v(u)=E_v = {\cal
B}(\Psi)$, and $\Psi(u)$ can be found explicitly:
\begin{equation}\label{84}
\Psi(u) = - Q \frac{E_v}{mc}r^{-1}(u) \,.
\end{equation}
On the other hand, the $\Psi(u)$ function happens to be a hidden
one, since in this degenerate case $U_{\alpha}$, $S_v$ and
$S_{\alpha}$ remain constant:
\begin{eqnarray}\label{85}
X_{\alpha}(u) = 0 \,, \quad U_{\alpha}(u) = C_{\alpha} \,,
\nonumber\\
S_{\alpha}(u) = \frac{E_v}{C_v} C_{\alpha} \,, \quad S_v(u) = E_v
= C_v E_0 \,.
\end{eqnarray}
The quantities $S_u(u)$ and $U_u(u)$
\begin{eqnarray}\label{86}
S_{u}(u) = -
\frac{E_v}{2C^2_v}[ 1 + g^{\alpha \beta}(u)C_{\alpha}C_{\beta}]\,,
\nonumber\\
U_{u}(u) =  \frac{1}{2C_v}[ 1 - g^{\alpha \beta}(u)
C_{\alpha} C_{\beta} ]
\end{eqnarray}
do not contain $\Psi(u)$ and consequently do not depend on $Q$.
This special type of particle motion is characterized by the
following feature: the projection of the polarization four-vector
onto the GW front plane is parallel to the projection of the
velocity four-vector not only at the initial moment, but during
the whole period of evolution, too.

\subsection{\small Second special solution: $E_v = 0$,\\ $Y_2(0) \cdot Y_3(0) =0$}

According to (\ref{52}), Eqs. (\ref{60}), (\ref{61}) describe an
exact solution to the evolution equations also in the case when $Q
\neq 0$, but $S_v(u) \equiv 0$. It is possible, when, first,
$S_v(0) = E_v = 0$, second, $X_2(0) \cdot X_3(0) = C^2_v E_2 E_3 =
0$, as can be seen from (\ref{53}) and (\ref{62}). In particular,
it is possible, when the initial polarization four-vector $S^i(0)$
has a vanishing projection onto the null Killing vector
$\mathop{\xi^i}\limits_{(v)}$ as well as onto one of the spacelike
Killing vectors $\mathop{\xi^i}\limits_{(2)}$ or
$\mathop{\xi^i}\limits_{(3)}$. The polarized boson with such
characteristics, traveling in the GW background, does not feel the
influence of the curvature-induced force. Both in first and the second
special cases, exact results of averaging over $Q$ coincide with
those for $Q \equiv 0$.

\subsection{\small Third special solution: $Y_2(0) = - Y_3(0) \neq 0$}
\label{III}

Eqs. (\ref{68}) show that if $Y_2(0) = - Y_3(0) \equiv Y(0)$,
then at an arbitrary retarded time instant
\begin{equation}\label{87}
Y_2(u) = - Y_3(u) = Y(0) \ {\rm e}^{- \Psi(u)} \,.
\end{equation}
Then Eq. (\ref{73}) takes the form
\begin{equation}\label{88}
\pm \int\limits_{0}^{\Psi(u)}
\frac{d\widetilde\Psi}{\sqrt{C^2_vE^2_0 - 2Y^2(0) {\rm
e}^{-2\widetilde\Psi} }} = -\frac Q{mc} r^{-1}(u) \,.
\end{equation}
Integration in (\ref{88}) yields
\begin{eqnarray}\label{89}
{\rm e}^{-\Psi(u)}=\frac{C_vE_0}{\sqrt{2} Y(0)}
\nonumber\\\times
\cosh^{-1}{\left[{\rm Arcosh}{\frac{C_vE_0}{\sqrt{2} Y(0)}}\mp
\frac{Q C_v E_0}{mc} r^{-1}(u)\right]} \,.
\end{eqnarray}
At the initial instant $u=0$, when $r^{-1}(0) =0$, one obtains
$\Psi(0) = 0$. Then at the instant $u^*$, when
\begin{equation}\label{90}
\frac{C_vE_0}{\sqrt{2} Y(0)} = \cosh{\frac{Q
C_vE_0}{mc}r^{-1}(u^*)} \,,
\end{equation}
the function ${\rm e}^{-\Psi}$ obviously reaches its maximum
value
\begin{equation}\label{91}
{\rm e}^{-\Psi(u^*)} = \frac{C_vE_0}{\sqrt{2} Y(0)} \,.
\end{equation}
The instant $u^*$ is a stopping point for the hyperbolic rotation
since the quantity $S_v(u^*)$ vanishes (see, (\ref{71})), and we
obtain $\Omega(u^*)=0$ and $\frac{d\Psi}{du}_{|u=u^*} = 0$. Using
(\ref{89}) in Eqs. (\ref{75})-(\ref{78}), we obtain an
example of an explicit representation of an exact solution
by elementary functions.

\subsection{\small Fourth special solution: $Y_2(0) = Y_3(0) \neq 0$}

Exact results for this case can be obtained from the results
of the Sec.~4.3. by the formal substitution $\Psi
\rightarrow - \Psi$.

\subsection{\small Statistical averaging in the general case}

Eqs. (\ref{71}) and (\ref{73}) show that, generally, the
dependence of $\Psi$ on the parameter $Q$ can be represented in
terms of elliptic function. For special initial data (see
Sec.~4.1-4.4) one can obtain the results in elementary
functions. Taking into account the relation (\ref{71}), one can
state, that the growth of $\Psi(u)$ is restricted by the
requirement $S_v^2 \geq 0$. When $S_v^2 = 0$, the function $\Psi$
reaches its extreme values
\begin{eqnarray}\label{92}
\exp\{2 \Psi_{1,2}\}
\equiv
 \frac{C^2_v E^2_0 \pm \sqrt{C^4_v
E^4_0 - (Y^2_2(0) - Y^2_3(0))^2 }}{(Y_2(0) + Y_3(0))^2} \,.
\end{eqnarray}
Both extreme values $\Psi_{1}$ and $\Psi_{2}$ correspond to the
relations
\begin{equation}\label{93}
\Omega(u^*) = 0 \,, \quad \left(\frac{d \Psi}{du}\right)_{|u=u^*}
= 0 \,.
\end{equation}
When $Y_2(0) \rightarrow - Y_3(0)$, the expression (\ref{92})
gives (\ref{91}), if we choose the minus sign in the numerator;
the plus sign gives the solution for the case $Y_2(0) \rightarrow
Y_3(0)$. Thus the function $\Psi(u)$ remains finite.

To perform averaging over $Q$, we consider the leading-order terms
in the expressions for $U_{\alpha}$ and $S_{\alpha}$. The
dimensionless parameter $(E_v/mc)r^{-1}(u)$ is considered to
be small. Since $E_v$ in quasiclassical theory is supposed to be
proportional to the Planck constant, the inequality
$(E_v/mc)r^{-1}(u) << 1$ does not require, in general, that
the GW is weak. When $Y_2(0)$ and $Y_3(0)$ are not  zero
simultaneously, the leading-order term in the expansion of the
function $\Psi(u)$ has exactly the same form as in (\ref{84}).
Averaging with the Gaussian function (\ref{83}) yields
\begin{eqnarray}\label{94}
\langle U_2\rangle_Q - U_{2_|Q=0}
\nonumber\\
={\cal D}^2\frac{ E_v^2 Y_3(0)}{4m^3c^3}\cdot\int\limits_0^u
\left\{{\rm e}^\beta \frac{(L^2\beta')'}{L}r^{-2} \right\}(\tilde
u)d\tilde u \,,
\end{eqnarray}
\begin{eqnarray}\label{95}
\langle U_3\rangle_Q - U_{3_|Q=0}
\nonumber\\
={\cal D}^2\frac{ E_v^2 Y_2(0)}{4m^3c^3}\cdot \int\limits_0^u
\left\{{\rm e}^{-\beta}
\frac{(L^2\beta')'}{L}r^{-2}\right\}(\tilde u)d\tilde u \,,
\end{eqnarray}
\begin{eqnarray}\label{96}
\langle S_v \rangle_Q - S_{v_|Q=0} = -\frac{{\cal
D}^2}{2E_vm^2c^2}\cdot r^{-2}(u)
\nonumber\\\times \{ 2 Y^2_2(0)Y^2_3(0) +
E_v^2[Y^2_2(0)+Y^2_3(0)]\} \,,
\end{eqnarray}
\begin{eqnarray}\label{97}
\langle S_2\rangle_Q-S_{2_|Q=0}=
\frac{{\cal D}^2Y_2(0)E_v^2}{4m^2c^2C_v}L{\rm e}^\beta r^{-2}(u)
\nonumber\\+
\frac{{\cal D}^2Y_3(0)E_v^3}{4m^3c^3C_v}
 \int\limits_0^u \left\{(L^2\beta')'\frac{{\rm e}^{\beta}}{L}r^{-2} \right\}(\tilde u) d\tilde u
\nonumber\\
-\frac{{\cal D}^2Y_2^2(0)Y_3(0)E_v}{m^3c^3C_v}r^{-1}(u)
 \int\limits_0^u \left\{(L^2\beta')'\frac{{\rm e}^{\beta}}{L}r^{-1} \right\}(\tilde u) d\tilde u
\nonumber\\
-\frac{{\cal D}^2\{2Y^2_2(0)Y^2_3(0)+E_v^2[Y_2^2(0)+Y_3^2(0)]\}}
{2m^2c^2 E_v C_v}
\nonumber\\\times
\left[C_2+\frac{Y_3(0)}{mc}(L{\rm }{\rm e}^\beta)'(u) \right]r^{-2}(u) \,,
\end{eqnarray}
\begin{eqnarray}\label{98}
\langle S_3\rangle_Q - S_{3_|Q=0}=
\frac{{\cal D}^2Y_3(0)E_v^2}{4m^2c^2C_v}Le^{-\beta} r^{-2}(u)
\nonumber\\+
\frac{{\cal D}^2Y_2(0)E_v^3}{4m^3c^3C_v} \int\limits_0^u \left\{
(L^2\beta')'\frac{{\rm e}^{-\beta}}{L}r^{-2} \right\}(\tilde u)
d\tilde u
\nonumber\\
-\frac{{\cal D}^2Y_2(0)Y_3^2(0)E_v}{m^3c^3C_v}r^{-1}(u)
 \int\limits_0^u \left\{(L^2\beta')'\frac{{\rm e}^{-\beta}}{L}r^{-1} \right\}(\tilde u) d\tilde u
\nonumber\\
-\frac{{\cal D}^2\{2Y^2_2(0)Y^2_3(0)+E_v^2[Y_2^2(0)+Y_3^2(0)]\}}
{2m^2c^2 E_v C_v}
\nonumber\\\times
\left[C_3-\frac{Y_2(0)}{mc}(L{\rm e}^{-\beta})'(u) \right]r^{-2}(u) \,.
\end{eqnarray}
We have used the notation
\begin{equation}\label{100}
\langle {\cal A } \rangle_Q \equiv \int^{\infty}_{-\infty} d Q
f(Q) {\cal A}(Q) \,.
\end{equation}
Thus the effect of random polarization-curvature force can lead
to essential changes in the particle state when the
dispersion parameter ${\cal D}$ is large enough.

\section{\normalsize Conclusions}

{\bf 1.}
We have formulated the master equations governing
the dynamics of a massive relativistic boson with an
arbitrary directed polarization four-vector, traveling
in the gravity field under the influence of an anomalous
polarization-curvature interaction (see Eqs. (\ref{33})
and (\ref{36})). The master equations constitute an analogue
of the Bargmann-Michel-Telegdi model \cite{BMT59}, describing
the evolution of an electron with an anomalous magnetic
moment. They use the direct analogy between the
electromagnetic and curvature interactions proposed by
Khriplovich \cite{KhP2000}, and represent a submodel of the Bander
and Yee general model \cite{BY93}.

{\bf 2.}
The master equations have been solved exactly for
the model of boson dynamics in a pp-wave gravitational
background. We have distinguished three exactly solvable
submodels. The first submodel describes the deterministic
anomalous polarization-curvature interaction
(see, (\ref{44}), (\ref{47}), (\ref{48}), (\ref{73}), (\ref{75})-(\ref{78})). The second one
represents its particular case with a vanishing coupling
constant of the anomalous polarization-curvature interaction
(see, (\ref{79})-(\ref{82})). The third submodel introduces a
random anomalous polarization-curvature interaction.

{\bf 3.}
The exact solutions of the master equations obtained
here demonstrate a specific type of the polarization four-vector
behaviour, designated as ``hyperbolic rotation" induced
by anomalous interaction of polarization with
the space-time curvature. ``Hyperbolic rotation" of the
polarization four-vector affects the boson dynamics; the
velocity four-vector depends non-linearly on the initial
data describing the direction of the polarization four-vector.

{\bf 4.}
The master equations happen to be nonlinear with
respect to the velocity four-vector $U^i$ as well as the
polarization four-vector $S^i$. This feature makes essentially
different the two models: the first one, where the
deterministic anomalous polarization-curvature interaction
vanishes, and the second one, where the random
coupling constant of the anomalous interaction vanishes
on the average. The corresponding distinction is demonstrated
by Eqs. (\ref{94})-(\ref{98}).

\Acknow
{The authors are grateful to W.~Zimdahl and V.G.~Bagrov for
fruitful discussions. The work was supported by the Russian
Foundation for Basic Research, the Russian Program of Support for the
Leading Scientific Schools (grant NSh-1789.2003.02) and the Deutsche
Forschungsgemeinschaft.}


\small

\begin{thebibliography}{99}

\bibitem{Abr03}
    M. Abraham, {\it Annalen der Physik\/} {\bf 10}, 105 (1903).

\bibitem{Thom26}
    L.H. Thomas, {\it Nature\/} {\bf 117}, 514 (1926).

\bibitem{Fre26}
    J. Frenkel, {\it Zs. Phys.\/} {\bf 37}, 243 (1926).

\bibitem{Tamm29}
    I. Tamm, {\it Zs. Phys.\/} {\bf 55}, 199 (1929).

\bibitem{Mat37}
    M. Mathisson, {\it Acta Phys. Polon.\/} {\bf 6}, 163 (1937).

\bibitem{Pap51}
    A. Papapetrou,
    {\it Proc. Roy. Soc. Lond.\/} {\bf A 209}, 248 (1951).

\bibitem{Dix64}
    W.G. Dixon, {\it Nuovo Cim.\/} {\bf 34}, 317 (1964).

\bibitem{BMT59}
    V. Bargmann, L. Michel and V.L. Telegdi,
    {\it Phys. Rev. Let.\/} {\bf 2}, 435 (1959).

\bibitem{Blo46}
    F. Bloch, {\it Phys. Rev.\/} {\bf 70}, 460 (1946).

\bibitem{Good62}
    R.H. Good (Jr.), {\it Phys. Rev.\/} {\bf 125}, 2112 (1962).

\bibitem{TBBD68}
    I.M. Ternov, V.G. Bagrov, V.A. Bordovitsyn and O.F. Dorofeev,
    {\it JETP\/} {\bf 55}, 2273 (1968) (in Russian).

\bibitem{BCL77-GF}
    A. Barducci, R. Casalbuoni and L. Lusanna,
    {\it Nucl. Phys.\/} {\bf B 124}, 521 (1977).

\bibitem{Hoj78}
    S. Hojman, {\it Phys. Rev.\/} {\bf D 18}, 2741 (1978).

\bibitem{BMSS80}
    A.P. Balachandran, G. Marmo, B.S. Skagerstam and A. Stern,
    {\it Phys. Lett.\/} {\bf B 89}, 199 (1980).

\bibitem{Rav80}
    F. Ravndal, {\it Phys. Rev.\/} {\bf D 21}, 2823 (1980).

\bibitem{CVZS81}
    G. Cognola, L. Vanzo, S. Zebrini and R. Soldati,
    {\it Phys. Lett.\/} {\bf B 104}, 67 (1981).

\bibitem{Bar82}
    A. Barducci, {\it Phys. Lett.\/} {\bf B 118}, 112 (1982).

\bibitem{Wos82}
    H.J. Wospakrik, {\it Phys. Rev.\/} {\bf D 26}, 523 (1982).

\bibitem{Aro88}
    H. Arod\'z, {\it Acta Phys. Polon.\/} {\bf B 19}, 697 (1988).

\bibitem{Khr89}
    I.B. Khriplovich,
    {\it JETP\/} {\bf 96}, 385 (1989) (in Russian).

\bibitem{vHo91}
    J.W. Van Holten, {\it Nucl. Phys.\/} {\bf B 356}, 3 (1991).

\bibitem{Yak92}
    M.Sh. Yakupov,
    {\it in:\/} ``Gravitatsija i Teorija Otnositelnosti'', N 29,
    Kazan Univ. Press, Kazan, 1992 (in Russian).

\bibitem{vHol93}
    J.W. Van Holten,
    ``Relativistic Dynamics of Spin in Strong External Fields''.
    hep-th/9303124.

\bibitem{BY93}
    M. Bander and K. Yee,
    {\it Phys. Rev.\/} {\bf D 48}, 2797 (1993).

\bibitem{CM94}
    J.P. Costella and B.H.J. McKellar,
    {\it Int. J. Mod. Phys.\/} {\bf A 9}, 461 (1994).

\bibitem{KhP96}
    I.B. Khriplovich and A.A. Pomeransky,
    {\it Phys. Lett.\/} {\bf A 216}, 7 (1996).

\bibitem{CGL97}
    M. Chaichian R., Gonzalez Felipe and D. Louis Martinez,
    {\it Phys. Lett.\/} {\bf A 236}, 188 (1997).

\bibitem{Gem97-2}
    G. Gemelli, {\it Gen. Rel. Grav.\/} {\bf 29}, 1163 (1997).

\bibitem{PSKh2000}
    A.A. Pomeransky, R.A. Sen'kov and I.B. Khriplovich,
    {\it UFN} {\bf 170}, 1129 (2000) (in Russian).

\bibitem{KhP2000}
    I.B. Khriplovich,
    ``Equations of Motion of Spinning Relativistic Particles''.
    hep-th/0009218.

\bibitem{KSHM-ESEE}
    D. Kramer, H. Stephani, M. McCallum, E. Herlt and E. Schmutzer,
    ``Exact Solutions of the Einstein Field Equations'',
    Deutcher Verlag der Wissenschaften, Berlin, 1980.

\bibitem{Moha2002}
    S. Mohanty, {\it Phys. Lett.\/} {\bf A 301}, 382 (2002).

\bibitem{BLM2003}
    A. Berard, J. Lages and H. Mohrbach,
    ``Classical Spinning Particle Assuming a Covariant Hamiltonian''.
    hep-th/0303189.

\bibitem{BBB-TRRP}
    V.G. Bagrov, G.S. Bisnovatyi-Kogan, V.A. Bordovitsyn et al.,
    ``Theory of Radiation of Relativistic Particles'',
    Fizmatlit, Moscow, 2002.

\bibitem{BK_FF_2001}
    A.B. Balakin and V.R. Kurbanova,
    {\it Foundations of Physics \/} {\bf 31}, 1039 (2001).

\bibitem{BK_SS_2001}
    A.B. Balakin and V.R. Kurbanova,
    {\it Spacetime and Substance\/} {\bf 2}, 82 (2001).

\bibitem{BKZ2002}
    A.B. Balakin, V.R. Kurbanova and W. Zimdahl,
    {\it Gravitation and Cosmology\/} {\bf 8}, 6 (2002).

\bibitem{BKZ2003}
    A.B. Balakin, V.R. Kurbanova and W. Zimdahl,
    {\it J. Math. Phys.\/} {\bf 44}, 5120 (2003).


\bibitem{Maugin}
    G.A. Maugin, {\it J. Math. Phys.\/} {\bf 19}, 1198 (1978).

\bibitem{Bal97}
    A.B. Balakin, {\it Class. Quantum Grav.\/} {\bf14}, 2881 (1997).

\bibitem{BK98}
    A.B. Balakin and V.R. Kurbanova,
    {\it in:\/} ``Recent problems in field theory. 1998'',
    ed. A.V. Aminova, Kantsler, Kazan, 1998.

\bibitem{Kur99}
    V.R. Kurbanova,
    {\it in:\/} ``Recent problems in field theory. 1999-2000'',
    ed. A.V. Aminova, Kazan, 2000.


\end{thebibliography}
\end{document}